\documentclass[thmsa,11pt,a4paper,oneside,final,ukenglish]{article}
\usepackage{graphicx}

\begin{document}

\title{Modeling the Role of Gap Junction Transport Characteristics in the Action
Potential Propagation}
\author{Isabel M. Irurzun,$^{1)}$, Magdalena M. Defeo,$%
^{2)}$\\
\\
(1) CCT La Plata- CONICET. Instituto de Investigaciones Fisicoqu\'{i}micas
Te\'{o}ricas y Aplicadas \\
(INIFTA), Facultad de Ciencias Exactas, Universidad Nacional de La Plata. \\
La Plata, Rep\'{u}blica Argentina. \\
(2) Hospital Interzonal General de Agudos ''Prof. Dr. RodolfoRossi''\\
La Plata, Rep\'{u}blica Argentina. \\
(*) Corresponding author: i\_irurzun@hotmail.com.ar}
\date{}
\maketitle
\begin{abstract}
We generalize the Cable Model to describe the transport characteristics
of the gap junctions coupling adjacent cells in the heart muscle.
Our model takes into account recent experimental information about
the time dependence of the junctional current and modifies the connections
between cells. It can be used with whatever excitable model is used
to represent the cell. We show that by modulating the gap junction
transport characteristics, it is possible to either suppress or produce
meandering of spiral waves, a state associated with cardiac arrhythmia.\newpage{}
\end{abstract}

\section{Introduction}

Spirals are generic structures in extended nonequilibrium systems.
They are typical of many reaction-diffusion systems, and have been
directly visualized in the heart muscle associated with reentrant
wave fronts {[}1{]}.

The spiral wave dynamics was numerically studied in both isotropic
and anisotropic media, and limited or unlimited geometries {[}2{]}-{[}6{]}.
If the spiral tip rotates around a stationary region, the activation
wave front will follow the same path from one complete reentrant cycle
to the next, and the associated ECG will exhibit a monomorphic pattern.
When the tip meanders, the activation sequence from one reentrant
cycle to the next will be different and the associated ECG can exhibit
polymorphic patterns. Polymorphic patterns appear in the ECG related
to ionic channel blockade, regions of refractory tissue, intracellular
calcium instabilities, etc {[}1{]}{[}7{]}-{[}10{]}.

More recently the challenge of the role of gap junctions in electrical
wave propagation has encourage investigations about the proarrhythmic
effects of reduced intercellular coupling{[}11{]}-{[}15{]}. Adjacent
cells are coupled by the myocardial gap junction channels, which transmit
the intercellular voltage gradients and allow the action potential
propagation. Disruption of the gap junction membrane structures terminates
the transfer of cardiac action potential across an electrically unexcitable
gap. Severe gap junctional uncoupling not only drastically reduces
conduction velocity, but also further results in meandering activation
wavefronts.{[}16{]},{[}17{]}

Simulations are usually made in the so-called Cable Model {[}2{]},{[}3{]}{[}18{]}-{[}20{]}
where the gap junction coupling is introduced as a constant conductance.
However, gap junctions are not passive but dynamic pores that allow
ions to pass from one cell to the next.

Experimental information also indicates that junctional current decays
exponentially when a constant voltage difference is applied across
the junction (pulse protocol).{[}21{]}-{[}26{]}

Other studies have considered the dynamic nature of the gap junctions\textbf{.
}Although these studies demostrated the significant dynamics that
can occur in the gap junction for the cases near decremental propagation
and conduction block, results did not deviate qualitatively from those
observed with the more classical representation of gap junctions as
constant resistors.{[}27{]}

\textit{In the present work, we generalize the Cable Model by introducing
the above-mentioned experimental finding and show that by modulating
the gap junction transport characteristics, it is possible either
to suppress or to produce meandering of spiral waves, a state associated
with cardiac arrhythmia.}

\section{The Model}

The origin of the action potential $V_{m}$ in an isolated cell lies
in the movement of ions throughout the cell membrane characterized
by a capacitance $C_{m}$. The process is described by

\begin{equation}
\frac{\partial V_{m}}{\partial t}=-\frac{I_{ion}(t)}{C_{m}}
\end{equation}
where $I_{ion}$ is the total ionic current, produced by the transport
of different ions throughout a number of ionic channels, and other
interchange mechanisms{[}28{]}. A value of $I_{ion}<0$ represents
the entrance of positive ions into the cell and an increase of $V_{m}$,
i.e., a depolarization of the cell. A reduction of $V_{m}$ (repolarization
of the cell) is produced by a value of $I_{ion}>0$, representing
the outgoing of positive ions from the cell. In general, $I_{ion}$
is the sum of outgoing and incoming currents ($I_{x}$), characterized
by different magnitudes and dependences on $V_{m}$

\begin{equation}
I_{ion}=\sum_{x=1}^{n}I_{x}
\end{equation}
A number of dynamic models have been used to derive $I_{ion}$.

The modification proposed in the present work affects the connections
between cells and therefore it can be applied whatever dynamic model
is used to express $I_{ion}$.

In the Cable Model the propagation of $V_{m}$ wave fronts is schematized
in Figure $1a$. An excitable model describes a membrane cell, and
several membrane cells are connected through resistors. The resistor
grid represents both the intracellular medium and the intercellular
channels, and the extracellular medium is assumed to have a negligible
resistance compared to the intracellular space.

The total cell current $I_{t}$ is given by

\begin{equation}
C_{m}\frac{\partial V_{m}}{\partial t}+I_{ion}=-I_{t}
\end{equation}
while the total intercellular current $I_{i}=I_{ix}+I_{iy}$ is given
by

\begin{equation}
g_{ix}\frac{\partial V_{m}}{\partial x}+g_{iy}\frac{\partial V_{m}}{\partial y}=I_{i}
\end{equation}
where $g_{ix}$ and $g_{iy}$ indicate both the intracellular and
the intercellular conductivities, and an anisotropic medium is assumed.
Note that $g_{ix}^{-1}$ and $g_{iy}^{-1}$ are extended resistivities,
i.e., $\left[g_{ix}^{-1}\right]=\left[g_{iy}^{-1}\right]=\Omega/m$.

Furthermore, the total cell current per unit length is given by

\begin{equation}
\frac{\partial I_{ix}}{\partial x}+\frac{\partial I_{iy}}{\partial y}\simeq\frac{\Delta I_{ix}}{\Delta x}+\frac{\Delta I_{iy}}{\Delta y}=-\left(\frac{I_{t1}}{\Delta x}+\frac{I_{t2}}{\Delta y}\right)=-\frac{I_{t}}{\Delta}
\end{equation}
where $\Delta x=\Delta y=\Delta$ is the unit length.

Therefore,

\begin{equation}
g_{ix}\frac{\partial^{2}V_{m}}{\partial x^{2}}+g_{iy}\frac{\partial^{2}V_{m}}{\partial y^{2}}=C_{m}\frac{\partial V_{m}}{\partial t}+I_{ion}
\end{equation}
where $\Delta$ has been omitted on the rigth-hand side for simplicity.

We note that the scale change 
\begin{eqnarray}
X & = & x\\
Y & = & \sqrt{\frac{g_{ix}}{g_{iy}}}y
\end{eqnarray}
eliminates the anisotropy effect in Eq. $6$. The quantity $g/C_{m}=D$
defines a diffusion coefficient.

But gap junction channels have time and $V_{i}$-dependent inactivation
properties that are dependent on the transjunctional or intercellular
voltage $V_{i}$. Experiments have been made on conexin 40 \ and
conexin 43 gap junctions, and the intercellular current $I_{i}$\ exponentially
decays in time with time constants depending on $V_{i}$.

In these experiments a constant transjunctional voltage $V_{i}$ is
applied using a pulse protocol where a $V_{i}$ pulse is repeated
five times and the ensemble average $I_{i}$ was fitted with an exponentially
decaying function\textbf{\ }to determine the decay time constants.
The reciprocal of the decaying time constants $(\tau)$ from $4$
to $10$ experiments at each $V_{i}$ were plotted fitted with the
general exponential expression {[}21{]}-{[}23{]}

\begin{equation}
\frac{1}{\tau}=A^{0}\exp(\left|V_{i}\right|/v_{0})
\end{equation}
where $A^{0}$and $v_{0}$ are constants\textbf{.}

\textit{These experimental findings indicate that the Cable Model
describing the intercellular channels as resistors must be improved
and time variations of }$I_{i}$\textit{ must be explicitly considered.}

The modification proposed in the present work is schematized in Figure
1b. 

\begin{figure}[h]
\begin{centering}
\includegraphics{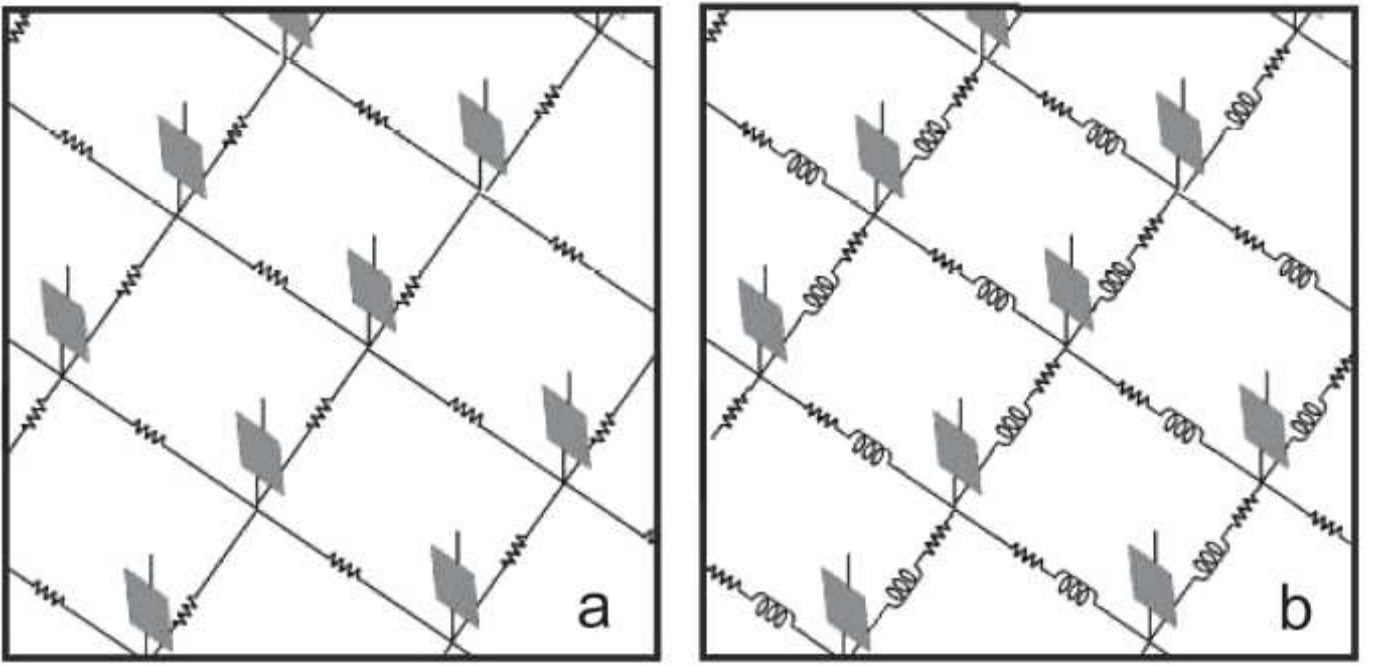} 
\par\end{centering}
\caption{(a) Cable Model scheme: each gray square represents an excitable model
describing the membrane cell, and several membrane cells are connected
through resistors; (b) Generalized Cable Model scheme: the membrane
cells are connected through resistors and inductances (recovery constant).}
\end{figure}

We consider the intercellular current $I_{i}$ as a recovery variable
and write Eq. 4 as

\begin{eqnarray}
\frac{\partial V_{m}}{\partial x} & = & \frac{1}{g_{ix}}I_{ix}+L_{ix}\frac{\partial I_{ix}}{\partial t}\\
\frac{\partial V_{m}}{\partial y} & = & \frac{1}{g_{iy}}I_{iy}+L_{iy}\frac{\partial I_{iy}}{\partial t}
\end{eqnarray}
where $L_{ix}$ and $L_{iy}$ are recovery constants and

\begin{equation}
\frac{1}{g_{ix}}=A_{0}\exp(\left|V_{ix}\right|/v_{c})
\end{equation}
with $A_{0}$ and $v_{c}$ constants, and

\begin{equation}
V_{ix}=\frac{\partial V_{m}}{\partial x}
\end{equation}
and identical equations in the $y$ direction.

Note that if $V_{ix}$ is constant $I_{i}$ is an exponentially decaying
function with a decay time constant depending of $V_{ix}$, as was
experimentally observed.

Assuming that $I_{ix}$ and $I_{iy}$ are analytical functions (i.e.,
$\frac{\partial}{\partial x}(\frac{\partial I_{ix}}{\partial t})=\frac{\partial}{\partial t}(\frac{\partial I_{iy}}{\partial x})$),
with the approximations $(g_{ix}v_{0})^{-1}<<1$, $(g_{iy}v_{0})^{-1}<<1$,
and Eq. 3 we obtain

\begin{equation}
C_{m}\frac{\partial V_{m}}{\partial t}+I_{ion}=-\left(I_{tx}+I_{ty}\right)
\end{equation}

\begin{eqnarray}
\frac{\partial I_{tx}}{\partial t}+\frac{1}{g_{ix}L_{ix}}I_{tx} & = & -\frac{1}{L_{ix}}\frac{\partial^{2}V_{m}}{\partial x^{2}}\\
\frac{\partial I_{ty}}{\partial t}+\frac{1}{g_{iy}L_{iy}}I_{ty} & = & -\frac{1}{L_{iy}}\frac{\partial^{2}V_{m}}{\partial y^{2}}\\
I_{t} & = & I_{tx}+I_{ty}
\end{eqnarray}
where $\Delta$ was omitted for simplicity on the lef-hand side of
Eqs. $15-16$, as in Eq. $6$. This Generalized Cable Model (GCM)
reduces to the former one (Eqs. $3$ and $4$) under the following
conditions:

(i) Stationary state for $I_{ix}$ and $I_{iy}$.

(ii) Isolated cells (i.e., $\frac{\partial^{2}V_{m}}{\partial x^{2}}=\frac{\partial^{2}V_{m}}{\partial y^{2}}=0$)
at $t\rightarrow\infty$($I_{ti}\rightarrow0$, $i=1,2$).

\section{Results}

We explored the GCM by using the cellular Barkley model to describe
the time variation of $I_{ion}$ as well as its dependences on $V_{m}${[}29{]}.
The Barkley model is perhaps one of the simplest cellular models showing
excitability.The cell is represented by an equivalent circuit containing
three elements connected in parallel: a capacitor representing the
cellular membrane, a variable resistor describing the ionic channels
and an inductance in series with a resistor representing the intracellular
medium. The Barkley model is extended by using the Cable Model with
a constant diffusion coefficient, and it is written as

\begin{eqnarray}
\frac{\partial u}{\partial t} & = & \varepsilon^{-1}u(1-u)\left[u-(v+b)/a\right]+D\nabla^{2}u\\
\frac{\partial v}{\partial t} & = & u-v
\end{eqnarray}
where $u$ and $v$ are the dimensionless versions of $V_{m}$ and
$I_{ion}$ respectively, $D$ is the diffusion coefficient, and $a$,
$b$ and $\varepsilon$ are parameters to model the nonlinear dependence
of $I_{ion}$ on $V_{m}$.

The space parameter of the Barkley model is shown in Figure 2 (modified
from Ref {[}29{]}). Stable 2D spiral waves are found in numerical
simulations inside the white region, and meandering of spiral waves
takes place in the MS region. The 2D medium does not support excitation
waves in region NW, is subexcitable in region SE, and is bistable
in region BI.

\begin{figure}[h]
\begin{centering}
\includegraphics{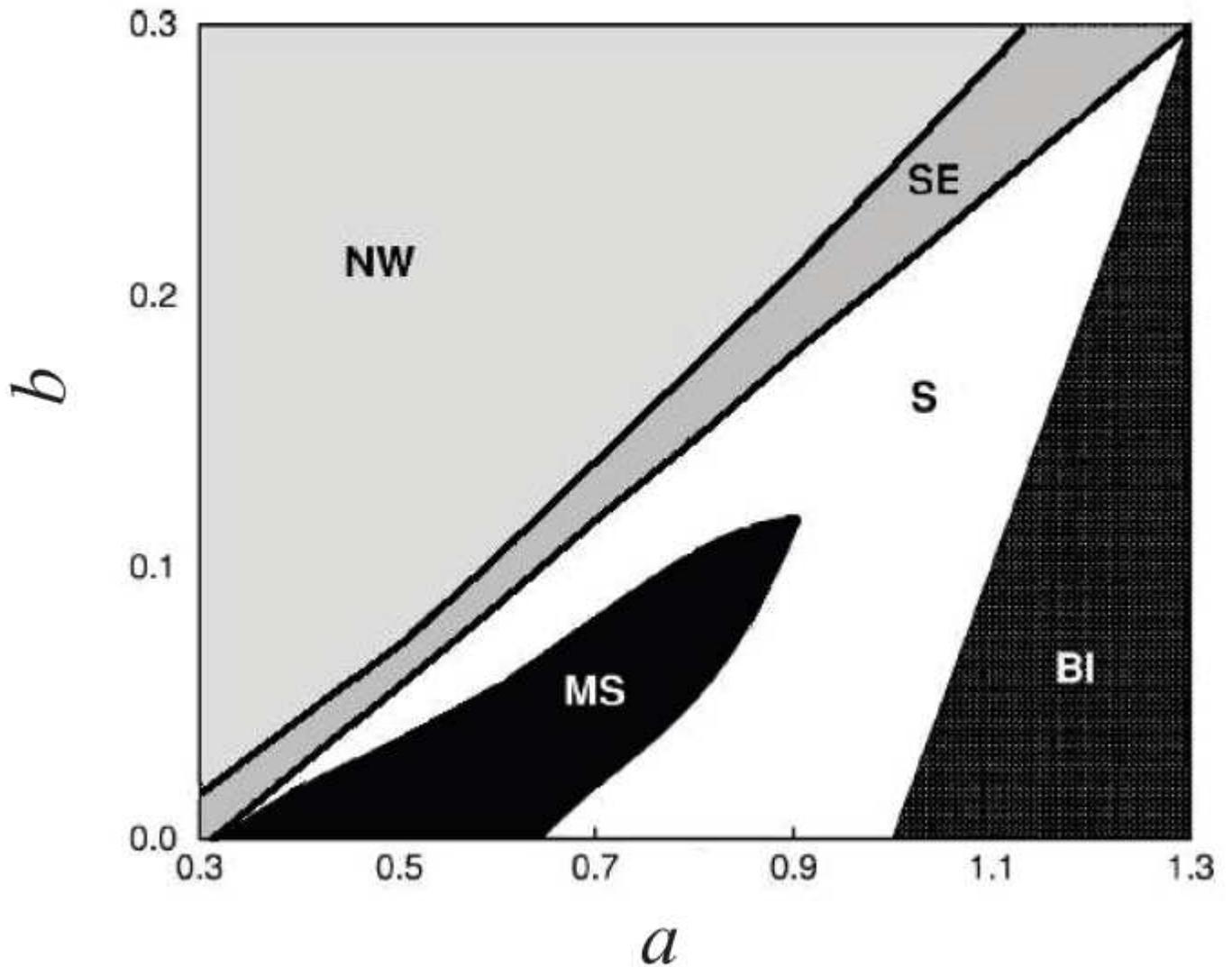} 
\par\end{centering}
\caption{Phase diagram of the spatially extended Barkley model (modified from
Ref: 29). MS: meandering of spirals, S: stable spirals, SE: subexcitable
zone, NW: no waves, BI: bistability.}
\end{figure}

In the present work we extend the Barkley model with the GCM and write

\begin{eqnarray}
\frac{\partial u}{\partial t} & = & \varepsilon^{-1}u(1-u)\left[u-(v+b)/a\right]-\sum_{j=1}^{2}w_{j}\\
\frac{\partial v}{\partial t} & = & u-v\\
\frac{\partial w_{j}}{\partial t} & = & -k_{j}w_{j}-k_{j}D\frac{\partial^{2}u}{\partial x_{j}^{2}}
\end{eqnarray}
where $w$ is the dimensionless version of $I_{t}$, and the subindex
$j=1,2$ considers the two Cartesian directions in a two-dimensional
tissue. Note that we do not explicitly reduce a cellular model to
its dimensionless form. This will depend on the model used and we
want to show only how two cells should be coupled taking into account
the basic properties of the gap junctions. We used the model constants
($A$ and $B$ in Eq. 23) to characterized its dynamical behaviour
in this new parametric space.

$k_{j}$ is given by

\begin{equation}
k_{j}=A\exp(u_{j}^{2}/B)
\end{equation}
with

\begin{equation}
u_{j}=\frac{\partial u}{\partial x_{j}}
\end{equation}
and where $A$ and $B$ are constants related with $A_{0}$ and $\nu_{c}$\textbf{. }

Note the anisotropy of $w_{j}$. Equation 23 takes into account the
experimentally observed dependence on $V_{i}$\ of the decay time
constants in the time dependence of $I_{i}$\textbf{\ }(see Eq. 9).
In this paper a quadratic dependence was introduced (instead of a
$\left|V_{i}\right|-$dependence) to preserve the even nature of the
function but to avoid the nondifferentiable point at $V_{i}=0$.

The model of Eqs. 20-22 reduces to the original Barkley model in the
following situations:

(i) Adiabatic conditions for $w_{j}$, i.e. $\frac{\partial w_{j}}{\partial t}=0.$

(ii) No diffusion of $u$ at $t\rightarrow\infty$ (i. e., $w_{j}\rightarrow0$).

In Eq. 23 the new parameters $A$ and $B$ regulate the transport
of different ions throughout the gap junction channels, and we show
that they determine the existence of either stable wave fronts (associated
with a normal electrical behavior of the heart) or meandering (associated
with arrhythmic behavior).

Figure 3 shows a characterization of our model in the $A-B$ parameter
space. Simulations were performed with $k_{j}D=1$, $\varepsilon=0.02$,
$a=0.7$, $b=0.05$. For these parameter values the original Barkley
model (with $D=1$) exhibits meandering of spirals. In the MS region
in Fig. 3 we obtain meandering of spirals, while in the S region there
are stable wave fronts. For $1/B=0$ (see Figure 3), the problem reduces
to one with constant diffusion $D=1/A$. In particular for $A=1$
($D=1$) we obtain stable wave fronts, differing from the original
Barkley model where meandering is obtained.

\textit{Therefore, the introduction of }$w_{j}$\textit{\ in the
GCM produces nontrivial modifications in the dynamic behavior of the
system.}

\begin{figure}[h]
\begin{centering}
\includegraphics{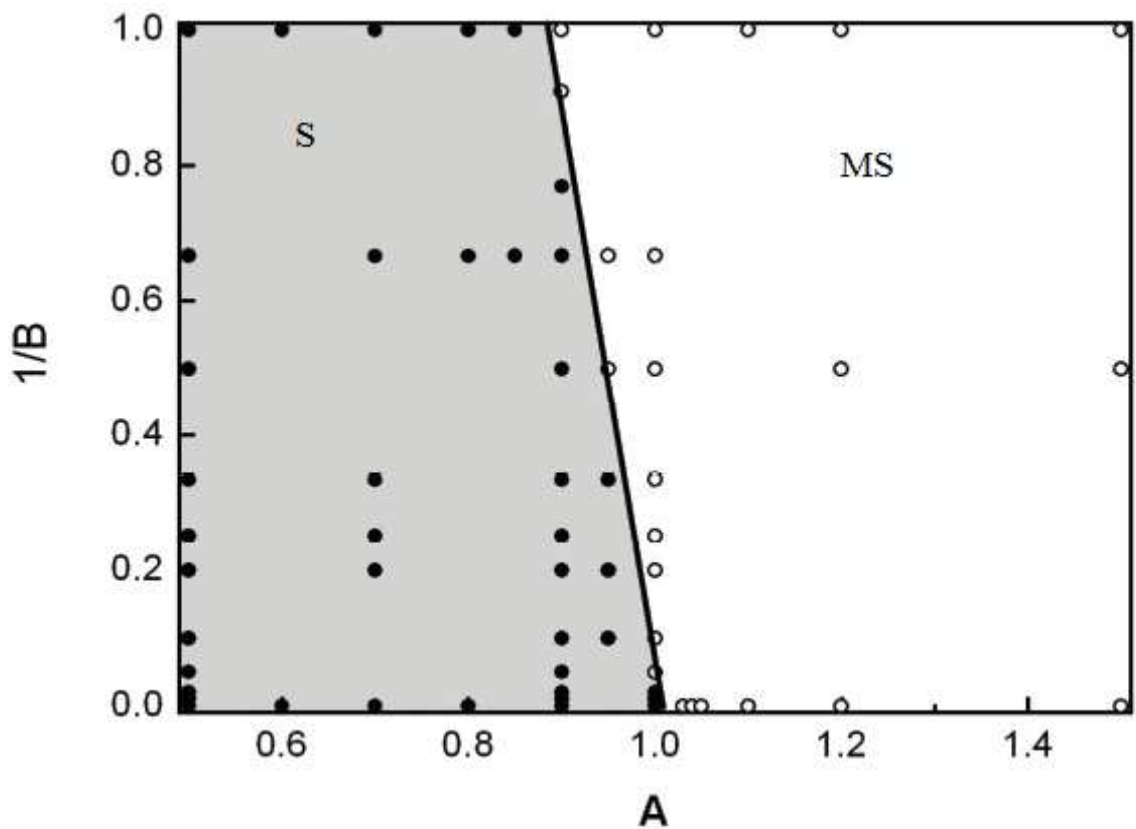} 
\par\end{centering}
\caption{Phase diagram of the modified spatially extended Barkley model. MS:
meandering of spirals, S: stable spirals. Simulations were performed
with $k_{j}D=1$, $\varepsilon=0.02$, $a=0.7$, $b=0.05$. }
\end{figure}

\section{Conclusions}

The results presented in this work show that the modification of the
gap junction transport properties can either suppress or favor the
development of spiral meandering.

Our conclusions are in agreement with experimental evidence showing
that severe gap junctional uncoupling results in meandering activation
wavefronts{[}16{]},{[}17{]}.

Of course the existence of meandering also depends on the cellular
model used in the simulation and it can be due to a number of other
mechanisms such as alterations in the excitability, anisotropy, etc.

The modelling of the gap junction transport properties performed in
this work explicitly introduces the exponential decaying of the intracellular
current $I_{t}$, by considering it as a recovery variable. In this
sense our model differs from other models that consider time and voltage
dependent conductances.{[}30{]},{[}31{]}

Alternatively, gap junctions can be considered as channels which are
''gated'' in a voltage-sensitive manner, so that the channels open
and close in response to the membrane potential. These gates can be
represented by using the Hodgkin-Huxley formalism, in which the conductance
term is decomposed into the product of a maximal conductance term
and one or more separate normalized variables that represent the probability
of finding the channel open. These variables follow their own differential
equations. The most common formulation for a gating variable $s_{i}$
is

\begin{equation}
\frac{\partial s_{i}}{\partial t}=(s_{0}-s)/ts
\end{equation}
where $s_{0}$ is the voltage dependent steady-state value of the
gate and $t_{s}$ is the voltage-dependent time constant of the gate.

This formalism applied to $I_{t}$ naturally leads to equations similar
to those used in the present work.

{\LARGE{}Acknowledgments}{\LARGE \par}

This work was supported by the Consejo Nacional y Ciencia y Tecnologís
(CONICET), Agencia Nacional de Promoción Científica y Tecnológica
(ANPCyT) and Universidad Nacional de La Plata (UNLP).\newpage{}

\end{document}